\documentstyle[11pt,moriond,epsfig]{article}

\def\Journal#1#2#3#4{{#1} {\bf #2}, #3 (#4)}

\def\NPB{{\em Nucl. Phys.} B}
\def\PLB{{\em Phys. Lett.}  B}
\def\PRL{\em Phys. Rev. Lett.}
\def\PRD{{\em Phys. Rev.} D}

\def\be{\begin{equation}}
\def\ee{\end{equation}}
\def\bea{\begin{eqnarray}}
\def\eea{\end{eqnarray}}

\begin{document}
\vspace*{4cm}
\title{PREHEATING THE UNIVERSE IN HYBRID INFLATION}

\author{JUAN GARC\'IA-BELLIDO}

\address{Theory Division, C.E.R.N., CH-1211 Gen\`eve 23, Switzerland}

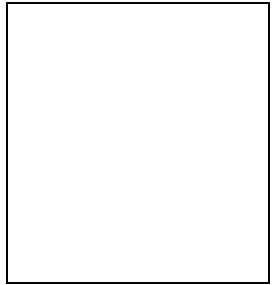
\maketitle 

\abstracts{One of the fundamental problems of modern cosmology is to
  explain the origin of all the matter and radiation in the Universe
  today. The inflationary model predicts that the oscillations of the
  scalar field at the end of inflation will convert the coherent
  energy density of the inflaton into a large number of particles,
  responsible for the present entropy of the Universe. The transition
  from the inflationary era to the radiation era was originally called
  reheating, and we now understand that it may consist of three
  different stages: preheating, in which the homogeneous inflaton
  field decays coherently into bosonic waves (scalars and/or vectors)
  with large occupation numbers; backreaction and rescattering, in
  which different energy bands get mixed; and finally decoherence and
  thermalization, in which those waves break up into particles that
  thermalize and acquire a black body spectrum at a certain
  temperature. These three stages are non-perturbative, non-linear and
  out of equilibrium, and we are just beginning to understand them. In
  this talk I will concentrate on the preheating part, putting
  emphasis on the differences between preheating in chaotic and in
  hybrid inflation. }

\section{Introduction}

At the end of inflation all the energy density is in the homogeneous
zero mode of the inflaton field. The Universe is in a vacuum-like
state with zero temperature and vanishing particle and entropy
densities. The problem of reheating is how to convert all this
coherent energy into a state of thermalized relativistic particles.
The original analysis~\cite{book} assumed the perturbative decay of
the inflaton into bosons and fermions, as if the inflaton were already
an ensemble of decoherent {\em particles}. Reheating ended when the
total decay rate of the inflaton was of the order of the expansion
rate of the Universe, $\Gamma\sim H$, while the total energy of the
inflaton field decayed exponentially fast into other particles. As a
consequence, the final reheating temperature only depended upon
$\Gamma$. We understand today that, for certain parameter ranges,
there is a new decay channel that is non-perturbative,\cite{KLS1}
due to the coherent oscillations of the inflaton field, which induces
stimulated emission of bosonic~\footnote{Fermions can also be
  parametrically amplified, but their occupation numbers are
  constrained, $n_k\leq1$, by Pauli's exclusion principle.} particles
into energy bands with large occupation numbers. The modes in these
bands can be understood as Bose condensates, and they behave like
classical waves. The backreaction of these modes on the homogeneous
inflaton field and the rescattering among themselves produce a state
that is far from thermal equilibrium and may induce very interesting
phenomena, such as non-thermal phase transitions~\cite{nonthermal}
with production of topological defects, a stochastic background of
gravitational waves,\cite{GW} production of heavy particles in a state
far from equilibrium, which may help GUT baryogenesis~\cite{baryo} or
constitute today the dark matter in our Universe.\cite{DM} These
classical waves eventually reach a state of turbulence where,
hopefully, decoherence will occur and thermalization will follow,
although these stages are not yet fully understood, either
analytically or numerically.

The period in which particles are produced via parametric resonance is
called {\em preheating}.\cite{KLS1} The idea is relatively simple, the
oscillations of the inflaton field induce mixing of positive and
negative frequencies in the quantum state of the field it couples to.
In the language of quantum fields in curved space, creation and
annihilation operators mix via Bogoliubov
transformations,\footnote{For an introduction to particle production
  in strong external fields, see Grib et al.\cite{GMM}} $a_k =
\alpha_k \bar a_k + \beta^\ast \bar a_{-k}^\dagger$, and with every
oscillation of the inflaton field, new particles are produced, $n_k
\equiv \langle\bar 0|a_k^\dagger a_k|\bar 0\rangle = |\beta_k|^2$. In
the case of chaotic inflation, with a massive inflaton $\phi$ coupled
to a massless scalar field $\chi$, the evolution equation for the
Fourier modes, $\ddot X_k + \omega_k^2 X_k=0$, with
$X_k=a^{3/2}(t)\chi_k$ and $\omega_k^2 = k^2/a^2(t) + g^2\phi^2(t)$,
can be cast in the form of a Mathieu equation, with coefficients
$A=k^2/4a^2m^2+2q$ and $q=g^2\Phi^2/4m^2$, where $\Phi$ is the
amplitude and $m$ is the frequency of inflaton oscillations,
$\phi(t)=\Phi(t)\sin mt$. For certain values of the parameters $(A,q)$
there are exact solutions that grow exponentially with time, and each
mode $k$ belongs to an instability band of the Mathieu
equation.\footnote{For a recent comprehensive review on preheating
  after chaotic inflation see Kofman~\cite{Kofman} and references
  therein.} These instabilities can be interpreted as coherent
``particle'' production with large occupation numbers. One way of
understanding this phenomenon is to consider the energy of these modes
as that of a harmonic oscillator, $E_k = |\dot X_k|^2/2 + \omega_k^2
|X_k|^2/2 = \hbar \omega_k (n_k + 1/2)$.  The occupation number of
level $k$ can grow exponentially fast, $n_k\sim\exp(2\mu_k mt)\gg1$,
and these modes soon behave like classical waves. It is analogous to
the well-known mechanism of generation of density perturbations during
inflation.\cite{book} The parameter $q$ during preheating determines
the strength of the resonance. It is possible that the model
parameters are such that parametric resonance does {\em not} occur,
and then the usual perturbative approach would follow, with decay rate
$\Gamma$. In fact, as the Universe expands, the growth of the scale
factor and the decrease of the amplitude of inflaton oscillations
shifts the values of $(A,q)$ along the stability/instability chart of
the Mathieu equation, going from broad resonance, for $q\gg1$, to
narrow resonance, $q\ll1$, and finally to the perturbative decay of
the inflaton.  Parametric resonance will stop whenever the inflaton
decay is dominated by the perturbative decay, $q m < \Gamma$, or when
the instability modes are redshifted away from the (last) narrow
resonance band, $q^2 m < H$.

\section{Preheating in hybrid inflation}

Until recently, preheating had been studied in chaotic and new
inflation only,\cite{KLS2} where the end of inflation occurs when the
rate of expansion is of order the mass of the field, $m\sim H$, and
the inflaton starts to oscillate around the minimum of its potential,
inducing particle production.\footnote{Although production of
  classical ``waves'' is more appropriate, we will nevertheless use
  the terminology of ``particle'' production with large occupation
  numbers.}  In a recent paper,\cite{GBL} we have studied the case of
hybrid inflation, where the end of inflation is triggered by the
symmetry breaking of another scalar field coupled to the inflaton, and
not by slow-roll.\cite{hybrid} This fact alone gives hybrid inflation
several advantages with respect to chaotic inflation. First, we can
consider models of inflation at low-energy scales, say the electroweak
scale, which nevertheless give the correct amplitude of temperature
fluctuations in the CMB. Second, we may have an effective frequency of
oscillation of one or both of the scalar fields being much larger than
the rate of expansion. As we will see, this induces a very efficient
and long-lived narrow resonance, where many oscillations occur in one
Hubble time, and a large number of particles are produced before their
momenta are redshifted by the expansion of the Universe. Preheating
can be very efficient in this case. Third, at the stage of
rescattering, gravitational waves are produced with wavelengths at
most of order the size of the horizon at that time, large enough today
to be detected at gravitational wave interferometers such as LIGO.

There are two fields in hybrid models, $\phi$ drives slow-roll
inflation and $\sigma$ triggers its end through a symmetry-breaking
potential,
\begin{equation}\label{hybrid}
V(\phi,\sigma) = {1\over4\lambda}(M^2-\lambda\sigma^2 )^2
 + {1\over2}m^2\phi^2 + {1\over2}g^2\phi^2\sigma^2 \,.
\end{equation}
These fields may couple to yet another one, $\chi$, with couplings
$(h_1^2\,\phi^2 + h_2^2\,\sigma^2)\chi^2/2$. Preheating depends very
strongly on the model parameters. During inflation, the
symmetry-breaking field $\sigma$ has a large mass, due to its coupling
to $\phi$, and is fixed at $\sigma=0$. The inflaton $\phi$ slow-rolls
down its effective potential $V(\phi)=V_0+m^2\phi^2/2$, driving
inflation and producing the metric fluctuations that later will give
rise to temperature anisotropies in the CMB and density perturbations
for large-scale structure. The amplitude of temperature anisotropies
on the scale of the horizon, as seen by COBE,\cite{COBE} constrains
the parameters of the model to satisfy
\begin{eqnarray}\label{COBE}
{g\over\lambda\sqrt\lambda}\,{M^5\over m^2M_{\rm P}^3} & \simeq & 
3.5\times10^{-5}\,,\\
n-1 \simeq  {\lambda\over\pi}\,{m^2M_{\rm P}^2\over M^4} & < & 0.2\,.
\end{eqnarray}
Contrary to the case of chaotic inflation, these constraints leave
plenty of freedom to chose the model parameters. One of the advantages
of hybrid inflation is that the rate of expansion at the end of
inflation, $H\sim M^2/\sqrt\lambda M_{\rm P}$, could be in a wide
range of scales, from just below Planck scale, as in models of
supergravity hybrid inflation,\cite{sugra} all the way to the
electroweak scale. In terms of the rate of expansion, COBE
constraints can be written as $g = 2\times 10^{-4} (n-1)\,M/H$ and
$n-1=2m^2/3H^2 < 0.2$.

When the inflaton $\phi$ falls below $\phi_c=M/g$, the field $\sigma$
triggers the end of inflation via a sudden or ``waterfall''
spontaneous symmetry breaking. This process occurs because of the
exponential growth of quantum fluctuations. In most models of hybrid
inflation, the mass-squared of the $\sigma$ field changes from large
and positive to large and negative in much less than one
$e$-fold.\footnote{There are hybrid models with two stages of
inflation where this condition is not satisfied.\cite{Guth,GBLW}} It
can then be shown that all tachyonic modes grow at approximately the
same speed and thus the $\sigma$ field can be described as a
homogeneous field.\cite{GBL} The behaviour of $\phi$ and $\sigma$
after the end of inflation therefore follows the homogeneous field
equations, and depends crucially on the ratio of couplings,
\begin{equation}\label{ratio}
{g\over\sqrt\lambda} \simeq (n-1)\,{M_{\rm P}\over M}\,10^{-4}\,.
\end{equation}
Depending on whether $g^2 \ll \lambda, \ g^2 \gg \lambda$ or $\,g^2
\sim \lambda$, the oscillations around the minimum will occur mainly
along the $\phi$ field, the $\sigma$ field, or both, respectively. Due
to their mutual couplings, the resonant production of particles could
be very efficient or completely suppressed, depending on this
ratio.\cite{GBL} Another important factor is the relation of the
frequency of oscillations to the rate of expansion. In chaotic
inflation this was fixed by the end of slow-roll condition, $m\sim
H$. However, in hybrid inflation we have the freedom to lower the rate
of expansion with respect to the mass $M$ and thus we can consider
models where the field oscillates many times, as much as $10^8$ times,
in one Hubble time.\cite{GBL} Thus, the resonant modes remain for a
long time in their instability bands and may produce large occupation
numbers before they are redshifted away by the
expansion.\cite{KLS2} In particular, in hybrid preheating, the narrow
resonance may be long-lived, $q^2 \bar m > H$, even for very small
parameters $q\ll1$, thanks to $\bar m\gg H$. This is a fundamental
difference with respect to chaotic inflation. Let us now consider in
some detail the three different cases mentioned above.

In the first case, $g^2 \ll \lambda$, the frequency of oscillations of
the $\phi$ field is $\bar m\simeq gM/\sqrt\lambda\gg m$ and, depending
on the model parameters, can be much greater than the rate of
expansion. For $\bar m \sim H$, the symmetry-breaking field soon
settles at its minimum $\sigma_0$, giving large masses to any field it
couples to, while the amplitude of inflaton oscillations quickly
decays with the expansion of the Universe. As a consequence, there is
essentially no production of either $\phi$ or $\sigma$
particles.\cite{GBL} However, if a massless scalar (or vector) field
$\chi$ couples to $\phi$, and not to $\sigma$ (for whatever symmetry
reason), it is possible to produce $\chi$-particles in the broad
resonance regime, for a wide range of couplings. This is the situation
that most resembles that of chaotic inflation; the growth parameter is
$\mu \sim 0.13$ and backreaction sets in after about 20 oscillations
of the $\phi$ field. In the other case, $\bar m \gg H$, the amplitude
of oscillations of $\phi$ after the end of inflation remains large, of
order $\phi_c$, during many oscillations, driving explosive particle
production even in the narrow resonance, for $\mu_k \ll 1$. Since the
rate of expansion is so small, the modes remain in the narrow
resonance for a long time, and backreaction on the inflaton
oscillations occurs before the modes are redshifted away from the
resonance.\cite{GBL} This is very different from the behaviour in
chaotic inflation. However, even if $\chi$ particles are quickly
produced, we still do not have a significant production of $\phi$ or
$\sigma$ particles before $\chi$-production backreacts on $\phi$.

In the case $g^2 \gg \lambda$, the situation described above is
reversed: it is now the symmetry-breaking field that oscillates around
its minimum, while the inflaton field becomes negligible. The
frequency of $\sigma$ oscillations is $\bar m = \sqrt2\,M$ and, as
before, it can be much larger than the rate of expansion. For $\bar m
\sim H$, the inflaton settles at $\phi=0$ while $\sigma$ oscillates
with ever-decreasing amplitude. In this case, there is an
insignificant production of either $\phi$ or $\sigma$ particles, due
to their large effective masses.\cite{GBL} Furthermore, if a field
$\chi$ couples to either $\phi$ or $\sigma$ its production via
parametric resonance is suppressed, because of the small amplitude of
$\phi$ oscillations or of the effective mass induced by $\sigma_0$,
respectively. It follows that preheating in this case is very
inefficient. On the other hand, for $\bar m \gg H$, explosive particle
production occurs for all fields, but mainly for $\sigma$ particles.
Even though the resonance is not particularly broad, $q\sim6$, the
growth parameter is very large, $\mu\sim0.3$, and there is very
efficient preheating in just a few oscillations, before backreaction
sets in. At the same time, production of $\phi$ and $\chi$ particles
is possible, although not as efficiently as for $\sigma$.\cite{GBL}
Note that preheating in this model could occur even in the absence of
extra fields $\chi$, simply due to the self-coupling of the
symmetry-breaking field.

Finally, in the case $g^2 \sim \lambda$, typical of certain models of
hybrid inflation in supergravity,\cite{sugra} the two homogeneous
fields oscillate with similar amplitudes and frequencies around the
global minimum $(\phi=0,\sigma=\sigma_0)$. Since they are coupled,
their frequencies and amplitudes vary rather chaotically and it takes
many oscillations for their behaviour to stabilize around a periodic
oscillation. During the chaotic motion, there can be no parametric
resonance since this effect requires a periodic behaviour. By the time
the oscillations become periodic, the amplitude has decreased so much
that not even the narrow resonance can be excited. Thus, for $\bar m
\sim H$, there is no particle production in either of the three
fields.\cite{GBL} However, for $\bar m \gg H$, there are many
oscillations in one Hubble time and eventually the motion becomes
periodic while the amplitude of oscillations is still large. This
results in a mild production of $\phi$ and $\sigma$ particles and an
explosive production of $\chi$ particles, if coupled only to the
$\phi$ field.\cite{GBL}

It should be noted that, in hybrid preheating, the equation that
describes the instability growth of modes $\chi_k$ is not exactly of
the Mathieu type. Even if the $\chi$ field couples only to one of the
fields, $\phi$ or $\sigma$, the fact that the two homogeneous fields
follow coupled equations implies that the frequencies of oscillations
of each field will vary strongly with time, unless the other field is
fixed at its minimum. This does no preclude parametric resonance, it
simply modifies the stability/instability chart. In some of the cases
we studied,\cite{GBL} the analysis with the Mathieu equation is a
reasonably good approximation; in others, one has to compute the
corresponding spectrum of instability bands. This is particularly
important for the case $g^2\sim\lambda$, where both fields oscillate
simultaneously in a rather chaotic way and their frequencies of
oscillation depend explicitly on the value of the other field, which
changes within one oscillation.

\section{Phenomenological consequences}

The processes by which the classical waves with large occupation
numbers produced at preheating decohere and become an ensemble of
relativistic particles in thermal equilibrium is still uncertain.
Numerical lattice simulations have been performed to try to understand
the process of backreaction and rescattering of those waves among
themselves and with the inflaton background.\cite{lattice} however, no
one has yet been able to compute the final reheating temperature of a
model in which preheating was important. It could be that, in the end,
the reheating temperature is precisely the one computed from the usual
perturbative analysis, $T_{\rm rh} \sim 0.1 \sqrt{\Gamma M_{\rm
P}}$. However, the reheating temperature is not the only observable
one can consider; in fact, preheating has opened the door to very
exotic phenomena that might have novel experimental signatures with
which to test our models of inflation. The interesting period is that
after backreaction and before thermalization, where large numbers of
particles (or waves) rescatter off themselves in a state far from
equilibrium. Among these new phenomena, there are non-thermal phase
transitions~\cite{nonthermal} with production of topological defects;
generation of gravitational waves;\cite{GW} or production of heavy
particles in a state far from equilibrium, which may help GUT
baryogenesis~\cite{baryo} or constitute the dark matter in our
Universe.\cite{DM}

I will concentrate here on the production of a stochastic background
of gravitational waves, which has specific signatures in the case of
hybrid preheating. The idea is the following: the collisions among
coherent waves of particles during rescattering radiates a reasonable
fraction of energy, typicaly $\sim 10^{-5}$, in the form of
gravitational waves. Their spectrum today depends on the details of
rescattering and can be computed only numerically in lattice
simulations.\cite{GW} However, the low-frequency end of the spectrum,
for wavelengths of order the size of the horizon at rescattering, can
be computed analytically since in that case it is dominated by the
gravitational bremsstrahlung associated with the scattering of $\chi$
particles off the inflaton condensate, with the corresponding
``evaporation'' of inflaton particles. Taking into account that the
occupation numbers of $\chi$ particles are typically of order
$n_k(\chi) \sim 10^2 h_1^{-2}$ at the end of rescattering,\cite{KLS2}
and assuming that decoherence and thermalization occurred immediately
after this stage, one can estimate the fraction of energy density in
gravitational waves today~\cite{GW}
\begin{equation}\label{GW}
\Omega_{\rm gw}(\omega)h^2 \sim \Omega_{\rm rad}h^2\,
{\bar m^2\over h_1^2M_{\rm P}^2}{\omega\over H_{\rm rs}}
\Big({g_0\over g_*}\Big)^{1/3}\,,
\end{equation}
where $\Omega_{\rm rad}h^2\simeq4\times10^{-5}$ and $\Omega_{\rm
  gw}(\omega) \equiv (\rho_c^{-1} d\rho_{\rm gw}/ d\ln\omega)_0$. On
the other hand, the wavelengths $2\pi/k_{\rm rs}$ of the gravitational
wave spectrum are redshifted to $\lambda\simeq 0.5(M_{\rm P}H_{\rm
  rs})^{1/2}k_{\rm rs}^{-1}$ today.\cite{GW} In particular, the
largest wavelength today corresponds to the size of the horizon at
rescattering, or $k_{\rm rs} \sim H_{\rm rs}$, and therefore the
minimum frequency in the spectrum is $f_{\rm min} \simeq
2\times10^2\,(H_{\rm rs}/{\rm TeV})^{1/2}$ Hz, which is right in the
centre, for $H_{\rm rs}\sim 1$ TeV, of the range of frequencies
detectable by LIGO, $10\,{\rm Hz}\leq f \leq 10^4$ Hz. In chaotic
inflation, the fraction in gravitational waves today is
typically~\cite{GW} of order $\Omega_{\rm gw} h^2\sim 10^{-12}$ at the
minimum frequency $f_{\rm min}\sim 10^6$ Hz, which lies outside the
range of frequencies and expected sensitivity of LIGO. The large rate
of expansion at rescattering turns out to be a disadvantage for
detecting gravitational waves from preheating in chaotic inflation.
One would require a much lower rate, of order a few TeV, to be within
the range of LIGO.

Let us consider now a concrete hybrid inflation model with a low rate
of expansion at the end of inflation, $H\sim10^2$ TeV. Such a model
could be consistent with COBE observations with very natural parameter
values, $g^2\sim0.01, \lambda\sim1, M \sim 10^{12}$ GeV and $m\sim1$
TeV. This model is of the first type, $g^2\ll\lambda$, and has a large
frequency of oscillations of the inflaton field, $\bar m\sim10^6 H$.
There will be explosive production of $\chi$ particles even in the
narrow resonance, $q \sim 10^4 h_1^2 \sim 10^{-3}$, with a very small
growth parameter $\mu\simeq q/2$. Backreaction in this model occurs at
$\bar m t_1 \simeq (1/4\mu) \ln(10^6\bar m(\bar m t_1)^3/h_1^5M_{\rm
  P})^{1/2} \sim 33/q$, which is still below one Hubble time, $t_1
\sim H^{-1}/30$, after the end of inflation. The occupation numbers of
$\chi$ particles at this stage can be estimated~\cite{KLS2} as
$n_k(\chi) \simeq 3\times10^2 h_1^{-2}q^{-1/4} \sim 30\,h_1^{-5/2}$.
Therefore, the estimate of Eq.(\ref{GW}) should be modified with an
extra factor $10h_1$ in the denominator. Taking $H_{\rm rs}\sim 10$
TeV at the end of rescattering, the fraction of energy density in
gravitational waves today is $\Omega_{\rm gw} h^2\sim 4\times10^{-10}$
at the minimum frequency $f_{\rm min}\sim 600$ Hz, just in the
appropriate range for detection by LIGO.

We conclude that preheating in hybrid inflation has many features that
differs from the usual chaotic or new inflation type of preheating,
and makes it phenomenologically attractive: inflation may occur at a
much lower scale, and still give the desired amplitude of temperature
anisotropies in the CMB; there could be many oscillations of the
inflaton field per Hubble time after inflation, which allows for a
long-lived narrow resonance and very efficient preheating; and it is
possible, in certain models, to generate a stochastic background of
gravitational waves in a range accesible to observations.

\section*{Acknowledgements}
I thank Andrei Linde for generous discussions and insights into the
problems of reheating.

\end{document}